\documentclass[prb,showpacs,preprintnumbers,amsmath,amssymb]{revtex4}  
\usepackage{epsf}
\usepackage{amsfonts}
\usepackage{amssymb}
\usepackage{amsthm}
\usepackage{graphicx}
\usepackage{hyperref}

\newcommand{\phdagger}{\mathop{\phantom{\dagger}}}           % phantom \dagger
\newcommand{\psiop}[1]{\psi^{\phdagger}_{#1}}                % psi annihilation
\newcommand{\psidop}[1]{\psi^{\dagger}_{#1}}                 % psi creation operator
\begin{document}

%\date{\today}
\title{Boundary Green's function for spin-incoherent interacting
  electrons in one dimension} 
\author{Paata Kakashvili$^1,^2$ and Henrik Johannesson$^3$}
\affiliation{$\mbox{}^1$Department of Applied Physics, Chalmers University of
Technology,
SE-412 96 G\"oteborg, Sweden}
\affiliation{$\mbox{}^2$Department of Physics \& Astronomy, Rice University, 6100 Main Street, Houston, TX 77005, USA}
\affiliation{$\mbox{}^3$Department of Physics, G\"oteborg University,
  SE-412 96 G\"oteborg, Sweden}

\begin{abstract}

The spin-incoherent regime of one-dimensional electrons has recently been 
explored using the Bethe ansatz and a bosonized path integral approach, revealing 
that the spin incoherence dramatically influences the correlations of charge
excitations. We here introduce a bosonization scheme for strongly
interacting electrons, allowing us to generalize the 
description to account for the presence of an open boundary. 
By calculating the single-electron Green's function we find that 
the charge sector power-law scaling is highly sensitive to the
boundary.
Our result allows for a detailed description of the
crossover between boundary and bulk regimes. We predict that scanning tunneling
microscopy on a spin-incoherent system will pick up oscillations
in the differential tunneling conductance as a function of the applied voltage $V$
at ``intermediate" distances $x$ from a real or a dynamically generated boundary.
The wavelength of the oscillations, $\pi v_c/x$, probes the speed $v_c$ of the charge 
excitations, and therefore the strength of the electron-electron interaction.

\end{abstract}


\maketitle

\section{introduction}\label{sec1}

The spin-incoherent regime of one-dimensional strongly interacting, very
low-density electrons has recently attracted a lot of interest~\cite{Fiete}. 
For zero 
temperature the kinetic energy of an electron can be estimated by the Fermi
energy $E_{F}=(\pi \hbar n)^{2}/8m$. For low densities, $n \ll
a_{B}^{-1}$, this energy is small compared to the Coulomb
potential energy $e^{2}n/\epsilon$ (with $\epsilon$ being the dielectric
constant and $a_{B}=\epsilon \hbar^{2}/me^{2}$ the effective Bohr
radius of the material). In the limit of low densities the system turns into
a \emph{Wigner crystal}~\cite{Schulz}, which $-$ in a classical picture $-$
can be viewed as a
system of electrons placed equidistantly so as to minimize the potential
energy. Quantum fluctuations induce an exponentially small
antiferromagnetic spin exchange $J>0$ between the electrons. In the
case when $J$ is the smallest energy 
scale, $J \ll T \ll E_{F}$, the spin exchange can no longer support 
collective spin excitations, and the system
is driven to a \emph{spin
incoherent} regime. The physics of the spin-incoherent regime has 
been addressed using the Bethe ansatz
\cite{CheianovZvonarev1,CheianovZvonarev2} and a bosonized 
path integral approach \cite{FieteBalents}. Surprisingly, it was found that
the spin incoherence dramatically influences the correlations of
charge excitations, leading to a power-law decay in the charge sector
with an interaction-dependent nonunitary exponent. 

In this paper we generalize the description in
Ref.~[\onlinecite{FieteBalents}] to account for the presence of an open
boundary. For this purpose we introduce a bosonization scheme, valid for strongly interacting electrons,
and presented in Sec.~\ref{sec2}.
In Sec.~\ref{sec3} we then derive the single-electron Green's
function in the presence of an open boundary condition. Scanning tunneling microscopy (STM) as a possible
experimental probe of the spin-incoherent regime is discussed in Sec.~\ref{sec4}.  The last section
contains a brief summary of our results. 

\section{Bosonization in the Strong Coupling Regime}\label{sec2}

We here introduce a bosonization scheme for strongly interacting electrons, 
for which the applicability of an ordinary bosonization 
\cite{GNT} becomes questionable. 
In the strong-coupling regime, the
electrons are localized at the lattice sites of the
Wigner crystal, and the usual procedure of linearizing the spectrum around
the Fermi points $\pm k_{F}$  and then expanding the electron fields in left and right
movers is no longer justified. In what follows we show that one can nonetheless perform an  
``effective" bosonization, valid for low energies and large distances. 

To begin with, it can be shown that a bosonic Hamiltonian correctly describes
the low-energy properties of a one-dimensional (1D) Wigner crystal~\cite{Matveev}.
To justify this statement we turn to the classical picture and 
describe a 1D Wigner crystal as a system of electrons vibrating around their 
equilibrium positions. This is very similar to the classical description of 
phonons as lattice vibrations. For low energies (long-wavelength limit) the 
vibrations can be described by elasticity theory, with the energy of the system expressed by the
Hamiltonian 
%%%%%%%%%%%%%%%%%%%%%%%%%%%%%%%%%%%%%%%%%%%%%%%%%%%%%%%%%%%%%%%%%%%%%%%
\begin{equation} \label{WignerClassical}
H=\int dx \left[ \frac{p^{2}}{2mn}+\frac{1}{2}
  mns^{2}(\partial_{x}u)^{2} \right].
\end{equation}
%%%%%%%%%%%%%%%%%%%%%%%%%%%%%%%%%%%%%%%%%%%%%%%%%%%%%%%%%%%%%%%%%%%%%%%%
Here $u(x)$ is the displacement of the medium and $p(x)$ is the
momentum density. Density fluctuations are given by $\delta n = -n
\partial_{x}u$, and  
%$s=\sqrt{(n/m)(\partial^{2}E/\partial  n^{2})}$,
$s$ plays the role of the speed of density waves.

The classical Hamiltonian in
Eq.~(\ref{WignerClassical}) can be straightforwardly quantized by imposing a
canonical commutation relation between $u(x)$ and $p(x)$,
%%%%%%%%%%%%%%%%%%%%%%%%%%%%%%%%%%%%%%%%%%%%%%%%%%%%%%%%%%%%%%%%%%%%%%%%%%
\begin{equation} \label{WignerCommutation}
[u(x),p(x')]=i\delta(x-x').
\end{equation}
%%%%%%%%%%%%%%%%%%%%%%%%%%%%%%%%%%%%%%%%%%%%%%%%%%%%%%%%%%%%%%%%%%%%%%%%

The resulting quantum Hamiltonian describes the propagation of density 
fluctuation in the Wigner crystal, and can be written in terms of a bosonic field
$\varphi_c(x)$ and its conjugate momentum $\Pi_c(x)$, connected to $u(x)$ and $p(x)$ by 
%%%%%%%%%%%%%%%%%%%%%%%%%%%%%%%%%%%%%%%%%%%%%%%%%%%%%%%%%%%%%%%%%%%%%%%%%%
\begin{equation} \label{DensityFluctuations}
u(x)=-\frac{\sqrt{2}}{\sqrt{\pi} n} \varphi_{c}(x), \hspace{1cm}
p(x)=-\frac{\sqrt{\pi} n}{\sqrt{2}} \Pi_{c}(x).
\end{equation}
%%%%%%%%%%%%%%%%%%%%%%%%%%%%%%%%%%%%%%%%%%%%%%%%%%%%%%%%%%%%%%%%%%%%%%%%
It follows that
%%%%%%%%%%%%%%%%%%%%%%%%%%%%%%%%%%%%%%%%%%%%%%%%%%%%%%%%%%%%%%%%%%%%%%%%%
\begin{equation} \label{BosonicHamiltonian}
H=\frac{v_{c}}{2}\int dx \left[ K_{c}
\Pi_{c}^{2}+\frac{1}{K_{c}}(\partial_{x}\varphi_{c})^{2}\right],
\end{equation}
%%%%%%%%%%%%%%%%%%%%%%%%%%%%%%%%%%%%%%%%%%%%%%%%%%%%%%%%%%%%%%%%%%%%%%%%%%
where
%%%%%%%%%%%%%%%%%%%%%%%%%%%%%%%%%%%%%%%%%%%%%%%%%%%%%%%%%%%%%%%%%%%%%%%%
\begin{equation}
v_{c}=s, \hspace{1cm}
K_{c}=\frac{v_{F}}{s},
\end{equation}
%%%%%%%%%%%%%%%%%%%%%%%%%%%%%%%%%%%%%%%%%%%%%%%%%%%%%%%%%%%%%%%%%%%%%%%%
with $v_{c}$ the speed of the propagating charge density fluctuations, and
with $v_{F}=\pi n/2m$ the Fermi velocity for noninteracting spinful
electrons.

We have thus managed to write the Hamiltonian for a 1D Wigner crystal in bosonized 
form although we did not know the bosonization formula which
connects electron and boson fields. We next derive such a formula, applying a kind of 
``reverse engineering" to the bosonic description of the Wigner crystal above. First we observe that in
the $J \rightarrow 0$ limit, electrons behave like \emph{spinless}
fermionic particles, i.e. only charge degrees of freedom survive and
spin degrees of freedom simply define the huge degeneracy of the ground
state. We may thus omit the spin index and write
%%%%%%%%%%%%%%%%%%%%%%%%%%%%%%%%%%%%%%%%%%%%%%%%%%%%%%%%%%%%%%%%%%%%%%%%%%%%%
\begin{equation}
\Psi(x)=\psiop{-}(x)+\psiop{+}(x),
\end{equation}
%%%%%%%%%%%%%%%%%%%%%%%%%%%%%%%%%%%%%%%%%%%%%%%%%%%%%%%%%%%%%%%%%%%%%%%%%%%%%
where $\psiop{-}(x)$ [$\psiop{+}(x)$] is the part of the electron operator which 
contains negative [positive] momenta. 
In analogy with ordinary bosonization of spinless fermions \cite{GNT} we introduce two
chiral boson fields $\phi_{c -}$ and $\phi_{c +}$, connected to $\varphi_c$ by
$\varphi_c = \phi_{c -} + \phi_{c +}$. We can then write an
effective bosonization formula, valid for low energies,
%%%%%%%%%%%%%%%%%%%%%%%%%%%%%%%%%%%%%%%%%%%%%%%%%%%%%%%%%%%%%%%%%%%%%%%%%%%%%%
\begin{eqnarray} \label{Bosonization1}
\psiop{\ell}(x) &\approx& \frac{1}{\sqrt{2 \pi \alpha}} e^{\ell i \tilde{k}_{F} x}
e^{\ell i \sqrt{\lambda} \phi_{c\ell}(x)}, \ \ \ \ell =\pm
\end{eqnarray}
%%%%%%%%%%%%%%%%%%%%%%%%%%%%%%%%%%%%%%%%%%%%%%%%%%%%%%%%%%%%%%%%%%%%%%%%%%%%%%
with $\alpha$ a short-distance cutoff, and where $\tilde{k}_{F}$ and $\lambda$
are to be determined. Using that the density operator is given by
%%%%%%%%%%%%%%%%%%%%%%%%%%%%%%%%%%%%%%%%%%%%%%%%%%%%%%%%%%%%%%%%%%%%%%%%%%%%%
\begin{equation} \label{FullDensity}
\rho=\Psi^{\dagger}(x)\Psi(x)=n+\sqrt{\frac{2}{\pi}}\partial_{x}\varphi_{c},    
\end{equation}
%%%%%%%%%%%%%%%%%%%%%%%%%%%%%%%%%%%%%%%%%%%%%%%%%%%%%%%%%%%%%%%%%%%%%%%%%%%%%
with the product of electron fields defined by point splitting \cite{GNT}, and
where $n$ is the average electron density and $(\sqrt{2/\pi})\partial_{x}\varphi_{c}$ 
is a fluctuation term [cf. Eqs. (\ref{WignerClassical}) and
(\ref{DensityFluctuations})], we find that
%%%%%%%%%%%%%%%%%%%%%%%%%%%%%%%%%%%%%%%%%%%%%%%%%%%%%%%%%%%%%%%%%%%%%%%%%%%%%%
\begin{eqnarray} \label{Parameters}
\tilde{k}_{F}=\pi n, \hspace{1cm} \lambda=8 \pi.
\end{eqnarray}
%%%%%%%%%%%%%%%%%%%%%%%%%%%%%%%%%%%%%%%%%%%%%%%%%%%%%%%%%%%%%%%%%%%%%%%%%%%%%%
We have here used that the fast oscillating nonchiral terms
$\psi^{\dagger}_{\pm}\psi_{\mp}$ contained in $\Psi^{\dagger}\Psi$
effectively average out to zero over large distances, and hence can be 
neglected in the long-wavelength limit. 
Also note that the doubling of the Fermi momentum in Eq. (\ref{Parameters}) 
is in agreement with our description of spinless
fermions, since now only a single fermion can occupy a given momentum state. 
As can be easily verified, the case of noninteracting spinless fermions
corresponds to $K_c=1/2$. It follows that in this bosonization scheme
interactions are absorbed in two steps: Local interactions among the spinful
electrons are incorporated in a free Hamiltonian for spinless fermions (with
$K_c=1/2$), while long-range interactions renormalize the value of $K_{c}$ (away
from $K_{c}=1/2$). To see this explicitly, note that for a delta function interaction 
among the electrons $-$ corresponding to the 
low-energy, long-wavelength limit of the infinite-$U$ 1D Hubbard model~\cite{Schulz}
$-$  the description as noninteracting spinless fermions becomes exact
(with $K_{c}=1/2$). 

To obtain a more conventional parametrization (where a unit value of an effective 
``charge parameter" $K$ corresponds to the case of noninteracting
spinless fermions) we define a bosonic field $\varphi \equiv \sqrt{2} \varphi_{c}$
with conjugate momentum $\Pi \equiv \Pi_c/\sqrt{2}$. Then the bosonization
formula in Eq. (\ref{Bosonization1}) and the Hamiltonian in Eq. (\ref{BosonicHamiltonian})
take the forms 
%%%%%%%%%%%%%%%%%%%%%%%%%%%%%%%%%%%%%%%%%%%%%%%%%%%%%%%%%%%%%%%%%%%%%%%%%%%%%%
\begin{eqnarray} \label{Bosonization2}
\psiop{\ell}(x) &\approx& \frac{1}{\sqrt{2 \pi \alpha}} e^{\ell i \tilde{k}_{F} x}
e^{\ell i \sqrt{4 \pi} \phi_{\ell}(x)}, \ \ \ \ell=\pm
\end{eqnarray}
%%%%%%%%%%%%%%%%%%%%%%%%%%%%%%%%%%%%%%%%%%%%%%%%%%%%%%%%%%%%%%%%%%%%%%%%%%%%%%
and
%%%%%%%%%%%%%%%%%%%%%%%%%%%%%%%%%%%%%%%%%%%%%%%%%%%%%%%%%%%%%%%%%%%%%%%%%
\begin{equation} \label{BosonicHamiltonian2}
H=\frac{v_{c}}{2}\int dx \left[ K
\Pi^{2}+\frac{1}{K}(\partial_{x}\varphi)^{2}\right],
\end{equation}
%%%%%%%%%%%%%%%%%%%%%%%%%%%%%%%%%%%%%%%%%%%%%%%%%%%%%%%%%%%%%%%%%%%%%%%%%%
respectively, with $K\equiv 2K_c=2 v_{F}/s= \tilde{v}_{F}/s$ (in agreement 
with the doubling of the Fermi momentum $\tilde{k}_{F})$.

\section{Boundary Green's Function}\label{sec3}

The bosonization procedure presented in the previous section can easily be adapted
to the case when a boundary is present. Imposing an open boundary condition 
(OBC) at the end, $x=0$, of a semi-infinite system with $x\ge 0$, we 
have that
%%%%%%%%%%%%%%%%%%%%%%%%%%%%%%%%%%%%%%%%%%%%%%%%%%%%%%%%%%%%%%%%%%%%%%%%%%%%%
\begin{equation}
\Psi(0)=\psiop{-}(0)+\psiop{+}(0)=0.
\end{equation}
%%%%%%%%%%%%%%%%%%%%%%%%%%%%%%%%%%%%%%%%%%%%%%%%%%%%%%%%%%%%%%%%%%%%%%%%%%%%%
Analytically continuing the chiral fermion operators to negative coordinates,
%%%%%%%%%%%%%%%%%%%%%%%%%%%%%%%%%%%%%%%%%%%%%%%%%%%%%%%%%%%%%%%%%%%%%%%%%%%%%
\begin{equation} \label{Continuation}
\psiop{+}(x)=-\psiop{-}(-x),
\end{equation}
%%%%%%%%%%%%%%%%%%%%%%%%%%%%%%%%%%%%%%%%%%%%%%%%%%%%%%%%%%%%%%%%%%%%%%%%%%%%%
and then following the standard procedure for an OBC~\cite{EggertAffleck}
$-$ using our Eq. (\ref{Bosonization2}) $-$ we obtain a bosonization formula
for strongly interacting spin-incoherent electrons with an OBC,
%%%%%%%%%%%%%%%%%%%%%%%%%%%%%%%%%%%%%%%%%%%%%%%%%%%%%%%%%%%%%%%%%%%%%%%%%%%%%
\begin{eqnarray} \label{BoundaryBosonization}
\psiop{-}(x) &\approx& \frac{1}{\sqrt{2 \pi \alpha}} e^{-i \tilde{k}_{F} x}
e^{-i \sqrt{4\pi}[\cosh(\vartheta)
\tilde{\phi}_{-}(x)-\sinh(\vartheta)\tilde{\phi}_{-}(-x)]}.
\end{eqnarray}
%%%%%%%%%%%%%%%%%%%%%%%%%%%%%%%%%%%%%%%%%%%%%%%%%%%%%%%%%%%%%%%%%%%%%%%%%%%%%
Here $e^{2 \vartheta}=K$, with $\tilde{\phi}_{-} = \cosh(\vartheta)\phi_-
- \sinh(\vartheta)\phi_+$ governed by the free chiral boson Hamiltonian
%%%%%%%%%%%%%%%%%%%%%%%%%%%%%%%%%%%%%%%%%%%%%%%%%%%%%%%%
\begin{equation} \label{ChiralBosonicHamiltonian}
H = \frac{v_c}{2}\int dx \,[\partial_x \tilde{\phi}_{-}(x)]^{2}.
\end{equation}
%%%%%%%%%%%%%%%%%%%%%%%%%%%%%%%%%%%%%%%%%%%%%%%%%%%%%%
To calculate the zero-temperature single-electron Green's function in the
presence of the OBC we follow the path integral approach introduced in
Ref.~[\onlinecite{FieteBalents}]. The averaging over spin introduces a factor
$2^{-N(x,x',\tau)}$ in the expression for the Green's function, where
$N(x,x',\tau)$ samples the number of electrons, or equivalently, the number of
noncrossing world lines in the interval $x-x'$. In the spin-incoherent regime,
here realized by first taking $J\rightarrow 0$ and then $T\rightarrow 0$, the
spin configurations all carry the same weight, and the probability that the
world lines in the interval $x-x'$ all have the same spin (as required for a
non-zero contribution to the low-energy Green's function) becomes equal to 
$2^{-N(x,x',\tau)}$. Given this, the calculation of the Green's function $G(x,x',\tau)$
translates into the calculation of four correlators for spinless time-boosted
fermions, with the spin averaging factor $2^{-N(x,x',\tau)}$ properly
inserted~\cite{footnoteA},
%%%%%%%%%%%%%%%%%%%%%%%%%%%%%%%%%%%%%%%%%%%%%%%%%%%%%%%%%%%%%%%%%%%%%%%%%%%%%
\begin{eqnarray}\label{GreensFunction}
G(x,x',\tau)&=&\langle 2^{-N(x,x',\tau)} \Psi(x,\tau) \Psi^{\dagger}(x',0)
\rangle \nonumber \\
&=&\langle 2^{-N(x,x',\tau)} \psiop{-}(x,\tau) \psidop{-}(x',0) \rangle+
\langle 2^{-N(x,x',\tau)} \psiop{+}(x,\tau) \psidop{+}(x',0) \rangle
\nonumber \\
&+&\langle 2^{-N(x,x',\tau)} \psiop{-}(x,\tau) \psidop{+}(x',0) \rangle+
\langle 2^{-N(x,x',\tau)} \psiop{+}(x,\tau) \psidop{-}(x',0) \rangle.
\end{eqnarray}
%%%%%%%%%%%%%%%%%%%%%%%%%%%%%%%%%%%%%%%%%%%%%%%%%%%%%%%%%%%%%%%%%%%%%%%%%%%%%
The operator $N(x,x',\tau)$ can be expressed as 
%%%%%%%%%%%%%%%%%%%%%%%%%%%%%%%%%%%%%%%%%%%%%%%%%%%%%%%%%%%%%%%%%%%%%%%%%%%%%
\begin{equation}
N(x,x',\tau)=n|x-x'|+\frac{1}{\sqrt{\pi}}[\varphi(x,\tau)-\varphi(x',0)],    
\end{equation}
%%%%%%%%%%%%%%%%%%%%%%%%%%%%%%%%%%%%%%%%%%%%%%%%%%%%%%%%%%%%%%%%%%%%%%%%%%%%%
where $n$ is the average electron (or world line) density, and with
$(1/\sqrt{\pi})[\varphi(x,\tau)-\varphi(x',0)]=[\cosh(\vartheta) +\sinh(\vartheta)]
[\tilde{\phi}_{-}(x,\tau)-\tilde{\phi}_{-}(-x,\tau) -\tilde{\phi}_{-}(x',0)
+ \tilde{\phi}_{-}(-x',0)]/\sqrt{\pi}$ the fluctuation term.

The correlators in Eq.~(\ref{GreensFunction}) are straightforwardly calculated
by first using Eq.~(\ref{Continuation}) to replace all occurrences of right-moving
fermion fields by left movers, and then applying the bosonization 
formula~(\ref{BoundaryBosonization}). Introducing relative and center-of-mass
coordinates, $r = x-x'$ and $R = (x+x')/2$, respectively, we obtain for the  
$G_{--}$ piece of the Green's function:
%%%%%%%%%%%%%%%%%%%%%%%%%%%%%%%%%%%%%%%%%%%%%%%%%%%%%%%%%%%%%%%%%%%%%%%%%%%%%
\begin{eqnarray} \label{LL}
G_{--}(x,x',\tau)&\equiv &\langle 2^{-N(x,x',\tau)} \psiop{-}(x,\tau)
\psidop{-}(x',0) \rangle \nonumber\\
&=&\frac{1}{2\pi \alpha}e^{-(\ln2/\pi) \tilde{k}_{F} |r|} e^{-i
  \tilde{k}_{F} r} e^{i \zeta_{--}} \frac{1}{(\alpha~\textrm{sign}\tau
  + v_{c} \tau + ir)}
\frac{1}{((\alpha~\textrm{sign}\tau + v_{c} \tau)^{2} + 
  r^{2})^{-2\Delta_1+2\Delta_2}} \nonumber \\
&\times&
\left(\frac{\sqrt{(\alpha^{2}+(2R+r)^{2})(\alpha^{2}+(2R-r)^{2})}}
{(\alpha~\textrm{sign}\tau + v_{c} \tau)^{2} + 
  4R^{2}}\right)^{2\Delta_1+2\Delta_3}\!\!\!\!\!\!\!\!\!\!\!\!\!\!\!\!\!\!\!, 
\end{eqnarray}
%%%%%%%%%%%%%%%%%%%%%%%%%%%%%%%%%%%%%%%%%%%%%%%%%%%%%%%%%%%%%%%%%%%%%%%%%%%%%
where
%%%%%%%%%%%%%%%%%%%%%%%%%%%%%%%%%%%%%%%%%%%%%%%%%%%%%%%%%%%%%%%%%%%%%%%%%%%%%
\begin{eqnarray}
\zeta_{--}&=&\frac{\ln 2}{4 \pi}K \left(\ln 
\frac{(\alpha^{2}+(2R+r)^{2})(\alpha^{2}+(2R-r)^{2})}
{((\alpha~\textrm{sign}\tau + v_{c} \tau)^{2} + 4R^{2})^{2}}
+ 2 \ln \frac{(\alpha~\textrm{sign}\tau + v_{c} \tau)^{2} + r^{2}}{\alpha^{2}}
\right) \nonumber \\
&+&\frac{\ln 2}{4 \pi} \left( \ln
\frac{(\alpha-i(r+2R))(\alpha-i(r-2R))}
{(\alpha+i(r+2R))(\alpha+i(r-2R))}
+2 \ln \frac{\alpha~\textrm{sign}\tau + v_{c} \tau + ir}
{\alpha~\textrm{sign}\tau + v_{c} \tau - ir} \right),
\end{eqnarray}
%%%%%%%%%%%%%%%%%%%%%%%%%%%%%%%%%%%%%%%%%%%%%%%%%%%%%%%%%%%%%%%%%%%%%%%%%%%%%
and with the exponents given by $\Delta_1=\frac{K}{8} \left( \frac{\ln 2}{\pi} \right)^{2}, \,
\Delta_2=\frac{1}{8} \left( \frac{1}{K}+K-2 \right), \,
\Delta_3=\frac{1}{8} \left( \frac{1}{K}-K \right).$

The $G_{-+}$ part of the Green's function reads
%%%%%%%%%%%%%%%%%%%%%%%%%%%%%%%%%%%%%%%%%%%%%%%%%%%%%%%%%%%%%%%%%%%%%%%%%%%%%
\begin{eqnarray} \label{LR}
G_{-+}(x,x',\tau)&\equiv &\langle 2^{-N(x,x',\tau)} \psiop{-}(x,\tau)
\psidop{+}(x',0) \rangle \nonumber \\
&=&\frac{1}{2\pi \alpha}e^{-(\ln2/\pi)\tilde{k}_{F} |r|} e^{-i 2\tilde{k}_{F} R}
e^{i \zeta_{-+}} \frac{1}{(\alpha~\textrm{sign} \tau+v_{c} \tau + 2iR)}
 \frac{1}{((\alpha~\textrm{sign} \tau + v_{c}\tau)^{2} + 4R^{2})
^{2\Delta_1+2\Delta_2}} \nonumber \\
&\times& \left(\sqrt{(\alpha^{2}+(2R+r)^{2})(\alpha^{2}+(2R-r)^{2})}
((\alpha~\textrm{sign}\tau
 + v_{c} \tau)^{2}  + r^{2}) \right)^{2\Delta_1} \nonumber \\
&\times&\left(\frac{\sqrt{(\alpha^{2}+(2R+r)^{2})(\alpha^{2}+(2R-r)^{2})}}
{(\alpha~\textrm{sign}\tau + v_{c} \tau)^{2} + 
  r^{2}}\right)^{2\Delta_3}\!\!\!\!\!\!\!\!, 
\end{eqnarray}
%%%%%%%%%%%%%%%%%%%%%%%%%%%%%%%%%%%%%%%%%%%%%%%%%%%%%%%%%%%%%%%%%%%%%%%%%%%%%
where
%%%%%%%%%%%%%%%%%%%%%%%%%%%%%%%%%%%%%%%%%%%%%%%%%%%%%%%%%%%%%%%%%%%%%%%%%%%%%
\begin{eqnarray} \label{LRPH}
\zeta_{-+}&=&\frac{\ln 2}{4 \pi} K
\ln\frac{\alpha^{2}+(2R+r)^{2}}{\alpha^{2}+(2R-r)^{2}} \nonumber \\
&+&\frac{\ln 2}{4 \pi} \left( \ln
\frac{(\alpha-i(r+2R))(\alpha-i(r-2R))}
{(\alpha+i(r+2R))(\alpha+i(r-2R))}
+2 \ln \frac{\alpha~\textrm{sign}\tau + v_{c} \tau + ir}
{\alpha~\textrm{sign}\tau + v_{c} \tau - ir} \right).
\end{eqnarray}
%%%%%%%%%%%%%%%%%%%%%%%%%%%%%%%%%%%%%%%%%%%%%%%%%%%%%%%%%%%%%%%%%%%%%%%%%%%%%
The expressions for the
$G_{++}$ and $G_{+-}$ pieces of the Green's function are immediately obtained
from Eqs. (\ref{LL}) - (\ref{LRPH}) by using that
%%%%%%%%%%%%%%%%%%%%%%%%%%%%%%%%%%%%%%%%%%%%%%%%%%%%%%%%%%%%%%%%%%%%%%%%%%%%%
\begin{eqnarray}
G_{++}(x,x',\tau)&=&G^{*}_{--}(x,x',\tau), \nonumber \\
G_{+-}(x,x',\tau)&=&G^{*}_{-+}(x,x',\tau).
\end{eqnarray}
%%%%%%%%%%%%%%%%%%%%%%%%%%%%%%%%%%%%%%%%%%%%%%%%%%%%%%%%%%%%%%%%%%%%%%%%%%%%%

Having derived the full expression for the boundary Green's function we can
study boundary and bulk regimes by taking the proper limits. The bulk regime
is defined by $R \gg v_c\tau,r$. For the chiral $G_{--}$ piece
we obtain
%%%%%%%%%%%%%%%%%%%%%%%%%%%%%%%%%%%%%%%%%%%%%%%%%%%%%%%%%%%%%%%%%%%%%%%%%%%%%
\begin{eqnarray}
G_{--}(x,x',\tau)&=&\langle 2^{-N(x,x'\tau)} \psiop{-}(x,\tau) \psidop{-}(x',0)
\rangle \nonumber\\
&=&\frac{1}{2\pi \alpha}e^{-(\ln2/\pi)\tilde{k}_{F} |r|} e^{-i
  \tilde{k}_{F} r} e^{i \zeta_{--}} \nonumber \\
&\times&\frac{1}{(\alpha~\textrm{sign}\tau
  + v_{c} \tau + ir)}
\frac{1}{((\alpha~\textrm{sign}\tau + v_{c} \tau)^{2} + 
  r^{2})^{-2\Delta_1+2\Delta_2}}~,
\end{eqnarray}
%%%%%%%%%%%%%%%%%%%%%%%%%%%%%%%%%%%%%%%%%%%%%%%%%%%%%%%%%%%%%%%%%%%%%%%%%%%%%\
in agreement with the result for the bulk Green's function derived in
Refs.~[\onlinecite{CheianovZvonarev1}] and [\onlinecite{FieteBalents}].
In contrast to the chiral part $G_{--}$ ($G_{++}$) of the Green's function,
$G_{-+}$ ($G_{+-}$) decays with a power law and vanishes in the bulk limit, as
it must: 
%%%%%%%%%%%%%%%%%%%%%%%%%%%%%%%%%%%%%%%%%%%%%%%%%%%%%%%%%%%%%%%%%%%%%%%%%%%%%
\begin{eqnarray}
G_{-+}(x,x',\tau) \sim \frac{1}{R^{K}}.
\end{eqnarray}
%%%%%%%%%%%%%%%%%%%%%%%%%%%%%%%%%%%%%%%%%%%%%%%%%%%%%%%%%%%%%%%%%%%%%%%%%%%%%

For the boundary case, defined by $v_{c} \tau >> r, R$, the chiral and nonchiral 
parts of the Green's function
show the same behavior, with the asymptotic scaling
%%%%%%%%%%%%%%%%%%%%%%%%%%%%%%%%%%%%%%%%%%%%%%%%%%%%%%%%%%%%%%%%%%%%%%%%%%%%%
\begin{equation}
G_{--}(x,x',\tau) \sim G_{-+}(x,x',\tau) \sim \frac{1}{\tau^{1/K}}.
\end{equation}
%%%%%%%%%%%%%%%%%%%%%%%%%%%%%%%%%%%%%%%%%%%%%%%%%%%%%%%%%%%%%%%%%%%%%%%%%%%%%
This agrees with the result in Ref.~[\onlinecite{FieteLeHur}] (see also
Ref.~[\onlinecite{Kindermann}]). Importantly, our general result
allows for a study of the crossover between boundary and bulk regimes.

\section{STM response in the spin-incoherent regime}\label{sec4}

In this section we discuss a possible experimental probe of the spin-incoherent
regime, using scanning tunneling microscopy (STM). Assuming that the STM tip
couples only to the conduction electrons, the differential tunneling conductance
$dI(V,x)/dV$ is directly proportional to the local electron tunneling density of states (LDOS)
$\rho(V,x)$, where $V$ is the applied voltage, and $x$ is the position of the tip as measured
from the boundary~\cite{TersoffHamann}. The boundary may here be one of the endpoints of a quantum wire
or a metallic nanotube, or generated dynamically \cite{KaneFisher} by an impurity
in the system (an antidot, or a deep ``core level" hole created by an X-ray 
photon \cite{FieteLeHur}). The LDOS is related to the boundary Green's function in Eq. (\ref{GreensFunction})
by
%%%%%%%%%%%%%%%%%%%%%%%%%%%%%%%%%%
\begin{equation}  \label{LDOS}
\rho(x,\omega) \approx -\frac{1}{\pi}\mbox{Im}G_R(x,\omega),
\end{equation}
%%%%%%%%%%%%%%%%%%%%%%%%%%%%%%%%%
where the retarded Green's function $G_R(x,\omega)$ is obtained
from the Fourier transform $G(x,\omega_n)$ of $G(x, x, \tau)$ in Eq.
(\ref{GreensFunction}) by analytically continuing $i\omega_n \rightarrow
\omega + i\eta_{+}$. The $\approx$ sign in Eq. (\ref{LDOS}) is a reminder 
that the expression for the Green's function in Eq. (\ref{GreensFunction})
as derived above becomes exact only in the long-wavelength limit
[cf. the text after Eq. (\ref{Parameters})].
The LDOS has a uniform [rapidly oscillating] part originating from the
chiral [nonchiral Friedel-type] terms $G_{--/++}(x,x,\tau)$ [$G_{-+/+-}(x,x,\tau)$]
in $G(x,x,\tau)$,
%%%%%%%%%%%%%%%%%%%%%%%%%%%%%%%%%%%
\begin{equation}  \label{N}
\rho(x,\omega) = \rho_{\mbox{\footnotesize{uni}}}(x,\omega) + \cos(2\tilde{k}_Fx)\rho_{\mbox{\footnotesize{osc}}}(x,\omega).
\end{equation}
%%%%%%%%%%%%%%%%%%%%%%%%%%%%%%%%%%
%%%%%%%%%%%%%%%%%%%%%%%%%%%%%%%%%%%%%%%%%%
\begin{figure}[!hpb]
\includegraphics[width=3in]{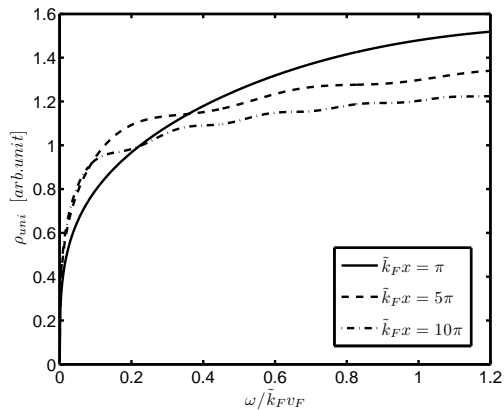} 
\caption{Uniform part of the LDOS, $\rho_{\mbox{\footnotesize{uni}}}$, as a function of energy $\omega$ for different
boundary-to-tip distances $x$. Oscillations for intermediate distances are
a result of interference of incoming and reflected charge modes at the boundary.}
\label{ldos} 
\end{figure}
%%%%%%%%%%%%%%%%%%%%%%%%%%%%%%%%%%%%%%%%%%

Figure 1 shows $\rho_{\mbox{\footnotesize{uni}}}$ for three choices of distance from the boundary,
obtained from Eq. (\ref{GreensFunction}) by numerically computing the
integrals which define the analytically continued Fourier transforms. 
The most interesting feature is the oscillation pattern seen for
``intermediate" distances where $\omega x \sim v_c$ (i.e., the crossover
regime between boundary and bulk behavior). This pattern is due to the
interference of incoming and reflected charge modes at the boundary.
The oscillations, of wavelength $\pi v_c/x$, give immediate information about 
the speed $v_c$ of the charge modes,
and therefore about the strength of the electron-electron interaction.
This unique signal of the spin-incoherent phase should be readily obtained
in STM experiments as an oscillation pattern in the differential tunneling conductance
as a function of the applied voltage $V$ (which fixes the frequency for which the
LDOS is probed). Let us here point out that the oscillations are present at {\em any}
distance from the boundary, but that in the boundary [bulk] regime, defined
by $\omega x \ll v_c$ [$\omega x \gg v_c$], the wavelength may become too long
[short] to be possible to resolve experimentally. (Cf. Fig. 1 where the wavelength
for the oscillation for the case $\tilde{k}_F x = \pi $ is roughly twice larger 
than the displayed frequency interval.)
It is important to stress that the Friedel terms [the last two terms in
Eq.~(\ref{GreensFunction})] do not change the result qualitatively. This is in
contrast to an ordinary Luttinger liquid where the interference of charge and
spin modes with different speeds produce an oscillation pattern with very different properties \cite{KJE}.
In the case of a spin-incoherent system there is only one propagating charge
mode with a single speed $v_c$, and the oscillation pattern of $\rho_{\mbox{\footnotesize{osc}}}$ 
will be the same as that of $\rho_{\mbox{\footnotesize{uni}}}$. 

Let us conclude this section by mentioning that other experimental
probes of spin-incoherence have been suggested in the literature, including momentum resolved
tunneling experiments on cleaved-edge overgrowth quantum wires \cite{FieteQian, Steinberg}, and 
quantum interference experiments on specially designed devices 
\cite{Kindermann}. It would be very interesting to explore how the characteristic crossover behavior
of the single-electron Green's function identified above may influence these proposed experiments. 

\section{Conclusions}\label{sec5}

We have calculated the exact single-electron boundary Green's function for
one-dimensional spin-incoherent electrons in the low-energy, long-wavelength
limit, using a bosonization scheme valid in the strong-coupling regime. The Green's
function thus obtained correctly reproduces known results in the bulk and
extreme boundary regimes. As
revealed by the expressions in Eqs. (\ref{LL}) and (\ref{LR}), the charge sector
power-law scaling of the Green's function is highly sensitive to the presence of
the boundary also at intermediate distances away from it, where the
center-of-mass coordinate $R \sim {\cal O}(v_c\tau)$. This feature will strongly
influence the local tunneling density of states, and may be probed experimentally
by scanning tunneling microscopy of one-dimensional conductors of spin-incoherent electrons.  

\begin{acknowledgments}
It is a pleasure to thank M.~Zvonarev for an inspiring discussion about
spin-incoherent electrons, and G.~Fiete for helpful communications. This paper
was supported by a grant from the Swedish Research Council.
\end{acknowledgments}


\begin{thebibliography}{99}
\bibitem{footnoteA} In contrast to Ref.~[\onlinecite{FieteBalents}] we do 
not include the permutation factor $(-1)^{N}$ in the expression for
the Green's function since it is automatically taken into account in
our spinless fermion formulation. In Ref.~[\onlinecite{FieteBalents}],
averages are carried out for spinless bosons, requiring the
permutation factor to be explicitly inserted (cf. the discussion 
after Eq.~(41) in Ref.~[\onlinecite{FieteQian}]).
\end{thebibliography}
\end{document}